\documentclass[graybox,envcountchap,sectrefs]{svmono}

\usepackage{mathptmx}
\usepackage{helvet}
\usepackage{courier}
\usepackage{type1cm}         
\usepackage{makeidx}         % allows index generation
\usepackage{graphicx}        % standard LaTeX graphics tool
\usepackage{multicol}        % used for the two-column index
\begin{document}

\author{\center Ricardo L\'opez-Ruiz \\ H\'ector Mancini \\ Xavier Calbet}
\title{\center A Statistical Measure of Complexity}
\subtitle{\center -- Book Chapter --}
\maketitle
%\date{July, 2010}
%\frontmatter%%%%%%%%%%%%%%%%%%%%%%%%%%%%%%%%%%%%%%%%%%%%%%%%%%%%%%

\tableofcontents
%\listoftables
%\listoffigures

%%%%%%%%%%%%%%%%%%%%% chapter.tex %%%%%%%%%%%%%%%%%%%%%%%%%%%%%%%%%
%
% sample chapter
%
% Use this file as a template for your own input.
%
%%%%%%%%%%%%%%%%%%%%%%%% Springer-Verlag %%%%%%%%%%%%%%%%%%%%%%%%%%

%\motto{}

\chapter{A Statistical Measure of Complexity}
\label{intro} % Always give a unique label
%\chaptermark{}
% to alter or adjust the chapter heading in the running head

%*******************************************************
%*********** PARTE DE INTRODUCCION *********************
%*******************************************************

\abstract{\emph{In this chapter, a statistical measure of complexity 
is introduced and some of its properties are discussed. 
Also, some straightforward applications are shown.}}

\newpage
\section{Shannon Information}

Entropy plays a crucial theoretical role in physics of macroscopic equilibrium
systems. The probability distribution of accessible states of a constrained
system in equilibrium can be found
by the inference principle of maximum entropy \cite{jaynes}.
The macroscopic magnitudes and the laws that relate them can be calculated
with this probability distribution by standard statistical mechanics techniques.

The same scheme could be thought for extended systems far from equilibrium,
but in this case we do have neither a method to find the
probability distribution nor the knowledge of the relevant magnitudes 
bringing the information that can predict the system's behavior.
It is not the case, for instance, with the metric properties 
of low dimensional chaotic systems by means of the Lyapunov exponents, 
invariant measures and fractal dimensions \cite{badii}.

Shannon information or entropy $H$ \cite{shannon49} can still be used as 
a magnitude in a general situation with $N$ accessible states:
\begin{equation}
H=-K\sum_{i=1}^{N}p_i\log{p_i}
\label{eq0}
\end{equation}
with $K$ a positive real constant and $p_i$ the normalized 
associated probabilities, $\sum_{i=1}^{N}p_i=1$.
An isolated system in equilibrium presents equiprobability, 
$p_i=1/N$ for all $i$, among its accessible states and 
this is the situation of maximal entropy,
\begin{equation}
H_{max}=K\log N.
\end{equation}
If the system is out of equilibrium, the entropy $H$ can be expanded
around this maximum $H_{max}$:
\begin{equation}
H(p_1,p_2,\ldots,p_N)=K\log N - \frac{NK}{2}\sum_{i=1}^{N}
\left(p_i-\frac{1}{N}\right)^2 +
\ldots = H_{max} - \frac{NK}{2} D + \ldots
\label{eq1}
\end{equation}
where the quantity $D=\sum_i (p_i-1/N)^2$, that we call {\it disequilibrium},
is a kind of distance from the actual system configuration to the equilibrium.
If the expression (\ref{eq1}) is multiplied by $H$ we obtain:
\begin{equation}
H^2 = H\cdot H_{max} - \frac{NK}{2}\; H\cdot D + K^2 f(N,p_i),
\end{equation}
where $f(N,p_i)$ is the entropy multiplied by the rest of the Taylor expansion 
terms, which present the form ${1\over N}\sum_i (Np_i-1)^m$ with $m>2$. 
If we rename $C=H\cdot D$,
\begin{equation}
C = cte\cdot H\cdot (H_{max} - H) +  K \bar{f}(N,p_i),
\label{eq5}
\end{equation}
with $cte^{-1}=NK/2$ and $\bar{f}=2f/N$. The idea of distance for 
the disequilibrium is now clearer if we see that  
$D$ is just the real distance $D\sim (H_{max}-H)$ for systems 
in the vicinity of the equiprobability.
In an ideal gas we have  $H\sim H_{max}$ and $D\sim 0$, then $C\sim 0$.
Contrarily, in a crystal $H\sim 0$ and $D\sim 1$, but also $C\sim 0$.
These two systems are considered as classical examples of simple models
and are extrema in a scale of disorder ($H$) or disequilibrium ($D$) but
those should present null complexity in a hypothetic measure of
{\it complexity}. This last asymptotic behavior is verified by the variable
$C$ (Fig. \ref{figIntro0}) and $C$ has been proposed as a  such magnitude \cite{lopezruiz95}.
We formalize this simple idea recalling the recent definition 
of {\it LMC complexity} in the next section.

Let us see another important property \cite{lopezruiz05} arising from relation (\ref{eq5}).
If we take the time derivative of $C$ in a neighborhood of equilibrium
by approaching $C\sim H(H_{max}-H)$, then we have
\begin{equation}
{dC\over dt} \sim -H_{max}{dH\over dt}.
\end{equation}
The irreversibility property of $H$ implies that ${dH\over dt}\geq 0$,
the equality occurring only for the equipartition,
therefore 
\begin{equation}
{dC\over dt} \leq 0.
\end{equation}
Hence, in the vicinity of $H_{max}$, LMC complexity is always decreasing on the evolution
path towards equilibrium, independently of the kind of transition and of the system 
under study. This does not forbid that complexity can increase when the system is very far 
from equilibrium. In fact this is the case in a  general situation as it can be seen,
for instance, in the gas system presented in Ref. \cite{calbet01}.

\section{A Statistical Complexity Measure}

On the most basic grounds, an object, a
procedure, or system is  said to be ``complex" when it does not match 
patterns regarded as simple. This sounds rather like an oxymoron 
but common knowledge tells us what is simple and complex: 
simplified systems or idealizations are always a starting point to solve
scientific problems. The notion of ``complexity"  
in physics  \cite{anderson91,parisi93}  starts
 by considering the perfect crystal and the 
isolated ideal gas as examples of simple models and therefore as systems 
with zero ``complexity".
Let us briefly recall their main characteristics with
``order", ``information" and ``equilibrium". 

A perfect crystal is completely ordered and the  
atoms are arranged  following stringent rules of symmetry. 
The probability distribution for the states accessible to the perfect
crystal is centered around a prevailing state of
perfect symmetry. A small piece of ``information" is enough to describe
the perfect crystal: the distances and 
the symmetries that define the elementary cell.
The ``information"  stored in this system can be considered minimal.
On the other hand, the isolated ideal gas is completely
disordered. The system  can be found in any of its accessible states with 
the same probability. All of them contribute in equal measure to
the ``information" stored in the ideal gas.
It has therefore a maximum ``information". These two  simple systems
are extrema in the scale of ``order" and ``information". It follows that
the definition of ``complexity" must not be made in terms of just ``order"
or ``information". 

It might seem reasonable to propose a measure of ``complexity"
by adopting some kind of distance from the equiprobable distribution of the
accessible states of the system \cite{lopezruiz95}. Defined in this way,
``disequilibrium"  would give an idea of
the probabilistic hierarchy of the system. 
``Disequilibrium" would be different from zero if there are privileged,
or more probable, states among those accessible. But this would not work.
Going back to the two examples we began with, it is readily seen that a perfect
crystal is far from an equidistribution among the accessible states
because one of them is totally prevailing, and so ``disequilibrium"
would be maximum. For the ideal gas, ``disequilibrium" would be zero by
construction. Therefore such a distance or ``disequilibrium" (a measure
of a probabilistic hierarchy) cannot be directly associated 
with ``complexity". 

\begin{figure}[h]  
\centerline{\includegraphics[width=12cm]{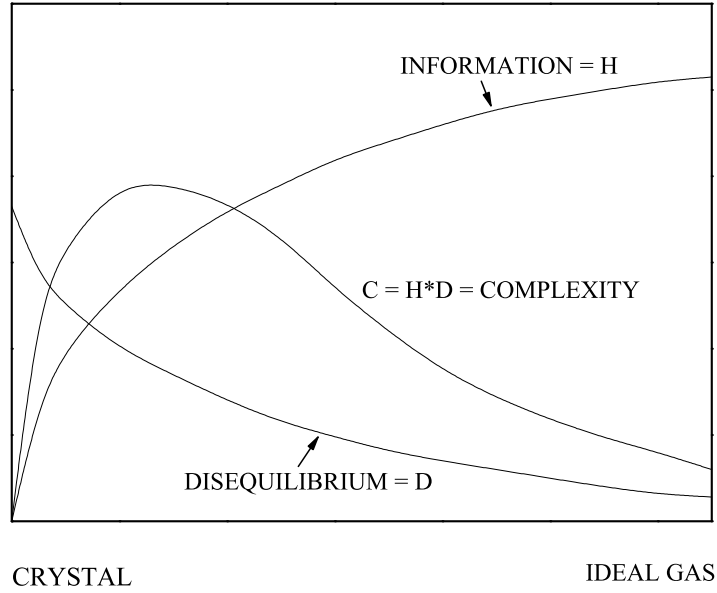}}  
\caption{Sketch of the intuitive notion of the magnitudes of 
``information" (H) and ``disequilibrium" (D) for the physical systems and
the behavior intuitively required for the magnitude 
``complexity". The quantity $C=H\cdot D$ is proposed to measure
such a magnitude.}  
\label{figIntro0}  
\end{figure}

In Figure \ref{figIntro0} we sketch an intuitive qualitative behavior for ``information"
$H$ and ``disequilibrium" $D$ for systems ranging from 
the perfect crystal to the ideal gas. As indicated in the former section,
this graph suggests that
the product of these two quantities  could be used 
as a measure of ``complexity": $C = H \cdot D$.
The function $C$ has indeed the features and asymptotic properties
that one would expect intuitively: it vanishes for the 
perfect crystal and for the
isolated ideal gas, and it is different from zero for the rest of the
systems of particles. We will follow these guidelines to establish a 
quantitative measure of ``complexity". 

Before attempting any further progress, however, we must recall that 
``complexity" cannot be measured univocally, because it depends on the
nature of the description (which always involves a reductionist process)
and on the scale of observation. Let us take an example to illustrate
this point. A computer chip can look very different at different scales.
It is an entangled array of electronic elements at microscopic scale but 
only an ordered set of pins attached to a black box at a macroscopic scale.

We shall now discuss a measure of ``complexity" based on the 
statistical description of systems.
Let us assume that the system has $N$ accessible states $\{x_1,x_2,...,x_N\}$ 
when observed at a given scale. We will call this an $N$-system. 
Our understanding of the behavior of this system determines the 
corresponding probabilities $\{p_1,p_2,...,p_N\}$ (with the 
condition $\sum_{i=1}^{N}p_i =1$) of each state ($p_i>0$ for all $i$). 
Then the knowledge of the underlying
physical laws at this scale is incorporated into a probability distribution 
for the accessible states. It is possible to find a quantity
measuring the amount of ``information". As presented in the former section,
under to the most elementary conditions of 
consistency, Shannon \cite{shannon49} determined the unique function 
$H(p_1,p_2,...,p_N)$ given by expression (\ref{eq0}),
that accounts for the ``information" stored in a system,
where $K$ is a positive constant. The  quantity $H$ is called 
{\it information}. The redefinition of information $H$ as some type of monotone function 
of the Shannon entropy can be also useful in many contexts. 
In the case of a crystal, a state $x_c$ would be the most probable
$p_c\sim 1$, and all others $x_i$ 
would be very improbable, $p_i\sim 0$ $i\neq c$. Then $H_c \sim 0$. 
On the other side, equiprobability
characterizes an isolated ideal gas, $p_i\sim 1/N$ so $H_g\sim K\log N$,
i.e., the maximum of information for a N-system.
(Notice that if one assumes equiprobability and $K=\kappa\equiv Boltzmann$ 
$constant$, $H$ is identified  with the thermodinamic
entropy, $S=\kappa\log N$). Any other N-system will have an amount of
information between those two extrema. 

Let us propose a definition of {\it disequilibrium} $D$ 
in a $N$-system \cite{prigo77}. The intuitive notion suggests that some kind of 
distance from an equiprobable distribution should be adopted.
Two requirements are imposed on the magnitude of $D$: $D>0$ in order to have a
positive measure of ``complexity" and $D=0$ on the limit of equiprobability.
The straightforward solution is to add the quadratic distances of 
each state to the equiprobability  as follows: 
\begin{equation}
D = \sum_{i=1}^{N}\left(p_i - \frac{1}{N}\right)^2\,.
\label{def-D}
\end{equation}
According to this definition, a crystal has maximum disequilibrium
(for the dominant state,
$p_c\sim 1$, and $D_c\rightarrow 1$ for $N\rightarrow \infty$)
while the disequilibrium for an 
ideal gas vanishes ($D_g\sim 0$) by construction. For any other system 
$D$ will have a value between these two extrema.

We now introduce the definition of {\it complexity} $C$ of 
a $N$-system \cite{lopezruiz95,lopezruiz94}.
This is simply the interplay between the information stored in 
the system and its disequilibrium:
\begin{equation}
C = H \cdot D = -\left ( K\sum_{i=1}^{N} p_i\log p_i \right ) \cdot
\left (\sum_{i=1}^{N}\left(p_i - \frac{1}{N}\right)^2 \right )\,.
\label{def-C}
\end{equation} 
This definition fits the intuitive arguments.
For a crystal, disequilibrium is large but the information stored
is vanishingly small, so $C\sim 0$.
On the other hand, $H$ is large for an ideal gas, but $D$ is small, 
so $C\sim 0$ as well. Any other system will have an 
intermediate behavior and therefore $C>0$.

As was intuitively suggested, the definition of complexity (\ref{def-C})
also depends on the {\it scale}.
At each scale of observation a new set of accessible states appears
with its corresponding probability distribution so that
complexity changes. Physical laws at each level
of observation allow us to infer the probability
distribution of the new set of accessible states, 
and therefore different values for $H$, $D$ and $C$ will be obtained. 
The straightforward passage to the case of a continuum number of states,
$x$, can be easily inferred. Thus we must treat with probability distributions
with a continuum support, $p(x)$, and normalization condition
$\int_{-\infty}^{+\infty}p(x)dx=1$. Disequilibrium has the limit
$D=\int_{-\infty}^{+\infty}p^2(x)dx$ and the complexity could be defined by:
\begin{equation} 
C=H\cdot D=-\left(K\int_{-\infty}^{+\infty}p(x)\log p(x)dx \right)
\cdot\left(\int_{-\infty}^{+\infty}p^2(x)dx \right)\,.
\label{def-C-continuo}
\end{equation} 
Other possibilities for the continuous extension of $C$
are also possible. For instance, a successful attempt of extending 
the LMC complexity for continuous systems
has been performed in Ref. \cite{catalan}. When the number of states available for
a system is a continuum then the natural representation is a continuous distribution.
In this case, the entropy can become negative.
The positivity of $C$ for every distribution is recovered by taking the exponential
of $H$ \cite{dembo91}. If we define $\hat{C}=\hat{H}\cdot D=e^{H}\cdot D$ as an extension of $C$
to the continuous case interesting properties characterizing the indicator $\hat{C}$ appear.
Namely, its invariance under translations, rescaling transformations and replication
convert $\hat{C}$ in a good candidate to be considered as an indicator bringing essential
information about the statistical properties of a continuous system.

\begin{figure}[h]  
\centerline{\includegraphics[width=8.5cm,angle=-90]{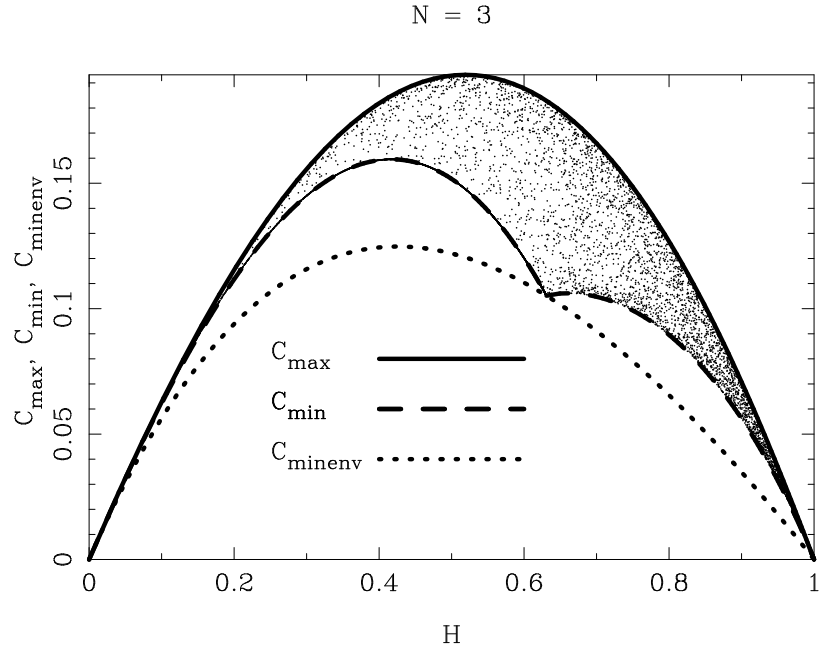}}  
\caption{In general, dependence of complexity ($C$)
on normalized information ($H$) is not univocal: many distributions $\{p_i\}$
can present the same value of $H$ but different $C$.
This is shown in the case $N=3$.}  
\label{figIntro1}  
\end{figure}  

\begin{figure}[h]  
\centerline{\includegraphics[width=7.5cm,angle=-90]{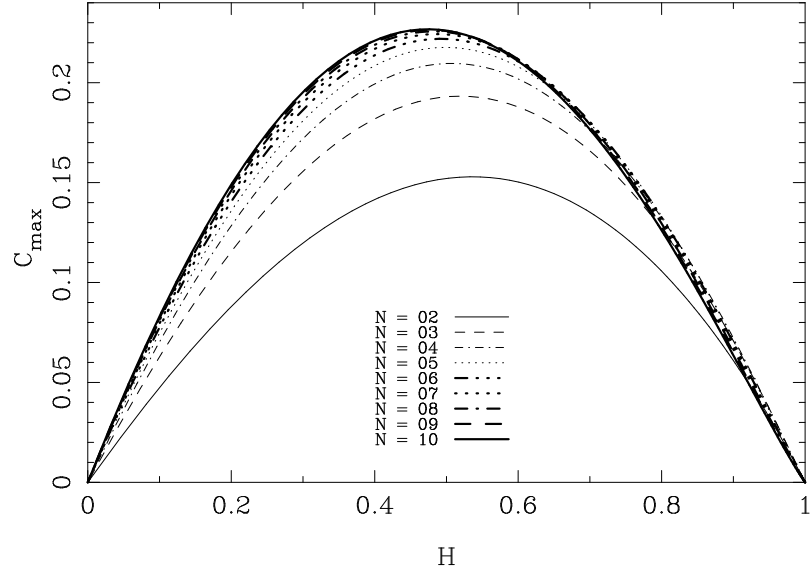}}  
\caption{Complexity ($C=H\cdot D$) as a function
of the normalized information ($H$) for a system with two accessible
states ($N=2$). Also curves of maximum complexity ($C_{max}$) 
are shown for the cases: $N=3,\ldots,10$.}  
\label{figIntro2}  
\end{figure}

Direct simulations of the definition give the values
of $C$ for general $N$-systems.
The set of all the possible distributions $\{p_1,p_2,...,p_N\}$ 
where an $N$-system could be found 
is sampled. For the sake of simplicity $H$ is normalized to the interval
$[0,1]$. Thus $H=\sum_{i=1}^{N} p_i\log p_i/\log N$.
 For each distribution $\{p_i\}$ the normalized information 
 $H(\{p_i\})$, and the disequilibrium 
$D(\{p_i\})$ (eq. \ref{def-D}) are calculated.
In each case the normalized complexity
$C=H\cdot D$ is obtained and the
pair $(H,C)$ stored.
These two magnitudes are plotted on a diagram $(H,C(H))$ in order
to verify the qualitative behavior predicted in Figure \ref{figIntro0}.
For $N=2$ an analytical expression for the curve 
$C(H)$ is obtained.
If the probability of one state is $p_1 =x$, that 
of the second one is simply $p_2 =1-x$. 
Complexity vanishes for the two simplest $2$-systems:
the crystal ($H=0$; $p_1 =1$, $p_2 =0$) and the ideal gas 
($H=1$; $p_1 =1/2$, $p_2 =1/2$). 
Let us notice that this curve is the simplest one that fulfills all
the conditions discussed in the introduction. 
The largest complexity is reached for $H\sim 1/2$ and its value
is: $C(x\sim 0.11)\sim 0.151$. 
For $N>2$ the relationship between 
$H$ and $C$ is not
univocal anymore. Many different distributions $\{p_i\}$ store
the same information $H$ but have different complexity 
$C$. Figure \ref{figIntro1} displays such a behavior for $N=3$. 
If we take the maximum complexity $C_{max}(H)$ associated with each 
$H$ a curve similar to the one for a $2$-system is recovered.
Every $3$-system will have a complexity below this line and upper the line
of $C_{min}(H)$ and also upper the minimum envelope complexity $C_{\rm minenv}$.
These lines will be analytically found in 
a next section. In Figure \ref{figIntro2} 
curves $C_{max}(H)$ for the cases 
$N=3,\ldots,10$ are also shown. 
Let us observe the shift of the complexity-curve peak 
to smaller values of entropy for rising $N$. This fact 
agrees with the intuition telling us that the biggest complexity
(number of possibilities of `complexification') be reached for lesser
entropies for the systems with bigger number of states.

Let us return to the point at which we started this discussion. 
Any notion of complexity in physics \cite{anderson91,parisi93} 
should only be made on the basis of a well defined or operational 
magnitude \cite{lopezruiz95,lopezruiz94}.
But two additional requirements are needed in order to obtain a 
good definition of complexity in physics: ($1$) the new magnitude
must be measurable in many different physical systems and ($2$) 
a comparative relationship and a physical interpretation 
between any two measurements should be possible.

Many different definitions of complexity have been proposed to date, mainly
in the realm of physical and computational sciences. Among these, several can be cited: 
algorithmic complexity (Kolmogorov-Chaitin) \cite{kolmogorov65,chaitin66},
the Lempel-Ziv complexity \cite{lempel76}, the logical depth of Bennett 
\cite{bennett85}, the effective measure complexity of
Grassberger \cite{grassberger86}, the complexity of a system based in its 
diversity \cite{huberman86}, the thermodynamical depth \cite{lloyd88},
the $\epsilon$-machine complexity \cite{crutchfield89} ,
the physical complexity of genomes \cite{adami00},
complexities of formal grammars, etc.
The definition of complexity (\ref{def-C}) proposed in this section offers 
a new point of view, based on a statistical
description of systems at a given {\it scale}. In this 
scheme, the knowledge of the physical laws governing the dynamic evolution 
in that scale is used to find its accessible states and its
probability distribution. This process would immediately indicate the
value of complexity. In essence this is nothing but an interplay
between the information stored by the system and the 
{\it distance from equipartition} (measure of a probabilistic hierarchy
between the observed parts) of the probability distribution of
its accessible states. 
Besides giving the main features of a ``intuitive" notion
of complexity, we will show in this chapter that we can go one step further 
and that it is possible to compute this quantity in relevant physical
situations \cite{calbet01,sanchez05,escalona09}. The most important point is 
that the new definition successfully enables us to discern situations regarded as complex.

\section{LMC Complexity: Extremal Distributions}

Now we proceed to calculate the distributions which maximize and minimize 
the LMC complexity and its asymptotic behavior \cite{calbet01}.

Let us assume that the system can be in one of its $N$ possible
accessible states, $i$. The probability of the system being in state $i$
will be given by the discrete distribution function, $f_i\ge 0$,
with the normalization condition $I \equiv \sum_{i=1}^N f_i = 1$.
The system is defined such that, if isolated, it will reach equilibrium, 
with all the states having equal probability, $f_{\rm e}=\frac{1}{N}$.
Since we are supposing that $H$ is normalized, $0 \leq H \leq 1$,
and $0 \leq D \leq (N-1)/N$, then complexity, $C$, is also normalized, $0 \leq C \leq 1$.

When an isolated system evolves with time, the complexity cannot have any possible 
value in a $C$ versus $H$ map as it can be seen in Fig. \ref{figIntro1}, but it
must stay within certain bounds, $C_{\rm max}$ and $C_{\rm min}$.
These are the maximum and minimum values of $C$ for a given $H$.
Since $C = D \cdot H$, finding the extrema of $C$ for constant $H$
is equivalent to finding the extrema of $D$.

There are two restrictions on $D$:
the normalization, $I$, and the fixed value of the entropy, $H$.
To find these extrema undetermined Lagrange multipliers are used. 
Differentiating expressions of $D$, $I$ and $H$, we obtain
\begin{eqnarray}
\frac{\partial D}{\partial f_j} & = & 2(f_j - f_{\rm e})\,,\\
\frac{\partial I}{\partial f_j} & = & 1\,,\\
\frac{\partial H}{\partial f_j} & = & -\frac{1}{\ln N}\left(\ln f_j + 1\right)\,.
\end{eqnarray}
Defining $\lambda_1$ and $\lambda_2$ as the Lagrange multipliers, we get:
\begin{equation}
2(f_j - f_{\rm e}) + \lambda_1 + \lambda_2(\ln f_j + 1)/\ln N = 0\,.
\end{equation}
Two new parameters, $\alpha$ and $\beta$, which are a
linear combinations of the Lagrange multipliers are defined:
\begin{equation}
f_j + \alpha \ln f_j + \beta = 0\,,
\label{eq:maxmin}
\end{equation}
where the solutions of this equation, $f_j$, are the values that minimize or
maximize the disequilibrium.

In the maximum complexity case
there are two solutions, $f_j$, to Eq.\ (\ref{eq:maxmin})
which are shown in Table \ref{tab:maximum}.
One of these solutions,
$f_{\rm max}$, is given by
\begin{equation}
\label{eq:defhmax}
H = - \frac{1}{\ln N} \left[ f_{\rm max} \ln f_{\rm max} + ( 1 - f_{\rm max} ) \ln
\left( \frac{1 - f_{\rm max}}{N - 1} \right) \right]\,,
\end{equation}
and the other solution by $(1 - f_{\rm max})/(N - 1)$.
The maximum disequilibrium, $D_{\rm max}$, for a fixed $H$ is
\begin{equation}
\label{eq:defdmax}
D_{\rm max} = (f_{\rm max} - f_{\rm e})^2 +
      (N - 1)\left(\frac{1 - f_{\rm max}}{N - 1} - f_{\rm e}\right)^2\,,
\end{equation}
and thus, the maximum complexity, which depends
only on $H$, is
\begin{equation}
\label{eq:defcmax}
C_{\rm max}(H) = D_{\rm max} \cdot H\,.
\end{equation}
The behavior of the maximum value of complexity versus $\ln N$ 
was computed in Ref. \cite{anteneodo96}.

\begin{table}
\caption{Probability values, $f_j$, that give a maximum
of disequilibrium, $D_{\rm max}$, for a given $H$.}
\label{tab:maximum}
\begin{center}
\begin{tabular}{|c|c|c|}
\hline 
Number of states & $ f_j$ & Range of $ f_j$ \\
with $ f_j$      &        &                 \\
\hline
\hline
$ 1$         & $f_{\rm max}$ & $ \frac{1}{N}\ \ldots\ 1$ \\
\hline
$ N - 1$     & $ \frac{1 - f_{\rm max}}{N - 1}$ & $ 0\ \ldots\ \frac{1}{N}$ \\
\hline
\end{tabular}
\end{center}
\end{table}

\begin{table}
\caption{Probability values, $f_j$, that give a minimum
of disequilibrium, $D_{\rm min}$, for a given $H$.}
\label{tab:minimum}
\begin{center}
\begin{tabular}{|c|c|c|}
\hline
Number of states & $ f_j$ & Range of $ f_j$ \\
with $ f_j$      &        &                 \\
\hline
\hline
$ n$         & $ 0$ & $ 0$ \\
\hline
$ 1$         & $ f_{\rm min}$ & $ 0\ \ldots\ \frac{1}{N - n}$ \\
\hline
$ N - n - 1$     &
         $ \frac{1 - f_{\rm min}}{N - n - 1}$ &
         $ \frac{1}{N - n}\ \ldots\ \frac{1}{N - n - 1}$\\
\hline
\end{tabular}
\end{center}
\begin{center}
{$n$ can have the values $0, 1,\ \ldots\, N-2$.}
\end{center}
\end{table}

\begin{figure}
\centerline{\includegraphics[width=8.5cm,angle=-90]{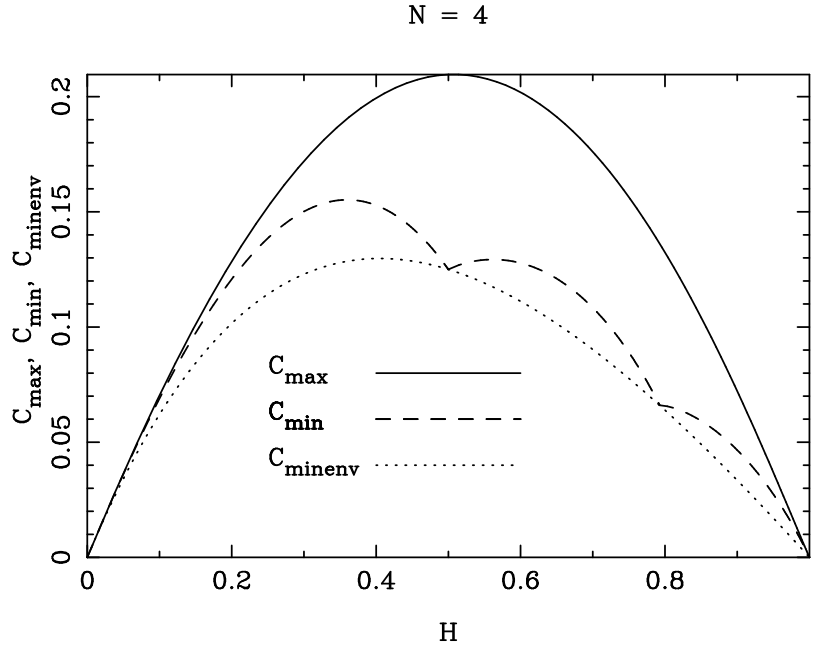}}
\caption{Maximum, minimum, and minimum envelope complexity, $C_{\rm max}$,
$C_{\rm min}$, and $C_{\rm minenv}$ respectively, as a function of
the entropy, $H$, for a system with $N=4$ accessible states.}
\label{fig:c_minmax}
\end{figure}

\begin{figure}
\centerline{\includegraphics[width=8.5cm,angle=-90]{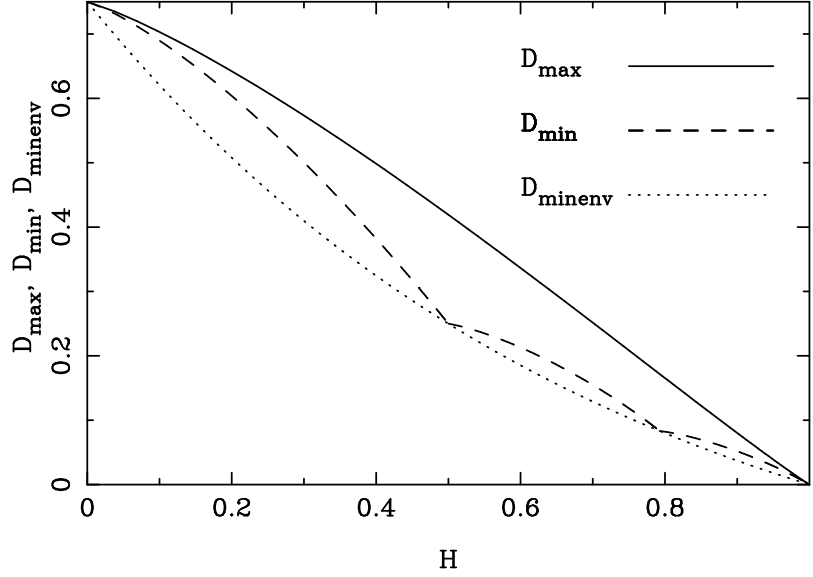}}
\caption{Maximum, minimum, and minimum envelope disequilibrium, $D_{\rm max}$,
$D_{\rm min}$, and $D_{\rm minenv}$ respectively, as a function of
the entropy, $H$, for a system with $N=4$ accessible states.}
\label{fig:d_minmax}
\end{figure}

Equivalently, the values, $f_j$, that give a minimum complexity
are shown in Table \ref{tab:minimum}. One of the solutions,
$f_{\rm min}$, is given by
\begin{equation}
H = - \frac{1}{\ln N} \left[ f_{\rm min} \ln f_{\rm min} + ( 1 - f_{\rm min} ) \ln
     \left( \frac{1 - f_{\rm min}}{N - n - 1} \right) \right]\,,
\end{equation}
where $n$ is the number of states with $f_j = 0$
and takes a value in the range $n = 0, 1,\ \ldots\ ,N - 2$.
The resulting minimum disequilibrium, $D_{\rm min}$, for a given $H$ is,
\begin{equation}
D_{\rm min} = (f_{\rm min} - f_{\rm e})^2 + (N - n - 1)
        \left(\frac{1-f_{\rm min}}{N - n - 1} - f_{\rm e}\right)^2 + n f_{\rm e}^2\,.
\end{equation}
Note that in this case $f_j = 0$  is an additional
hidden solution that stems from the positive restriction
in the $f_i$ values. To obtain these solutions explicitly we can define $x_i$ such that
$f_i \equiv {x_i}^2$. These $x_i$ values do not have the restriction of positivity 
imposed to $f_i$ and can take a positive or negative value.
If we repeat the Lagrange multiplier method with these
new variables a new solution arises: $x_j = 0$, or equivalently, $f_j = 0$.
The resulting minimum complexity, which again only depends on $H$, is
\begin{equation}
\label{defcmin}
C_{\rm min}(H)= D_{\rm min} \cdot H\,.
\end{equation}
As an example, the maximum and minimum of complexity, $C_{\rm max}$ and $C_{\rm min}$,
are plotted as a function of the entropy, $H$, in Fig.~\ref{fig:c_minmax}
for $N=4$. Also, in this figure, it is shown the minimum envelope complexity,
$C_{\rm minenv}=D_{\rm minenv}\cdot H$, where $D_{\rm minenv}$ is defined below.
In Fig.~\ref{fig:d_minmax} the maximum and minimum disequilibrium,
$D_{\rm max}$ and $D_{\rm min}$, versus $H$ are also shown.

As shown in Fig.~\ref{fig:d_minmax} the minimum disequilibrium function is piecewise
defined, having several points where its derivative
is discontinuous. Each of these function pieces corresponds to a different value of
$n$ (Table \ref{tab:minimum}).In some circumstances it might be helpful
to work with the ``envelope'' of the minimum disequilibrium
function. The function, $D_{\rm minenv}$, that traverses
all the discontinuous derivative points in the $D_{\rm min}$ versus $H$ plot is
\begin{equation}
D_{\rm minenv} = e^{- H \ln N} - \frac{1}{N}\,,
\end{equation}
and is also shown in Figure~\ref{fig:d_minmax}. 

When $N$ tends toward infinity the probability, $f_{\rm max}$,
of the dominant state has a linear dependence with the entropy,
\begin{equation}
\lim_{N \rightarrow \infty} f_{\rm max} = 1 - H\,,
\end{equation}
and thus the maximum disequilibrium scales as 
$\lim_{N \rightarrow \infty} D_{\rm max} = (1-H)^2$.
The maximum complexity tends to
\begin{equation}
\label{eq:cmaxlim}
\lim_{N \rightarrow \infty} C_{\rm max} = H \cdot (1-H)^2\,.
\end{equation}
The limit of the minimum disequilibrium and complexity vanishes,
$\lim_{N \rightarrow \infty} D_{\rm minenv} = 0$, and thus
\begin{equation}
\label{eq:cminlim}
\lim_{N \rightarrow \infty} C_{\rm min} = 0\,.
\end{equation}
In general, in the limit $N \rightarrow \infty$,
the complexity is not a trivial function of the entropy,
in the sense that for a given $H$ there exists
a range of complexities between $0$ and $C_{\rm max}$,
given by Eqs. (\ref{eq:cminlim}) and (\ref{eq:cmaxlim}), respectively.

In particular, in this asymptotic limit, the maximum of $C_{\rm max}$ is found when
$H=1/3$, or equivalently $f_{\rm max}=2/3$,
which gives a maximum of the maximum complexity of $C_{\rm max}=4/27$.
This value was numerically calculated in Ref. \cite{anteneodo96}.

\section{R\'enyi Entropies and LMC Complexity}

Generalized entropies were introduced by R\'enyi \cite{renyi} in the form of
\begin{equation}
I_q={1\over 1-q}\log\left(\sum_{i=1}^N p_i^q\right),
\end{equation}
where $q$ is an index running over  all the integer values. By differentiating $I_q$ with 
respect to $q$ a negative quantity is obtained independently of $q$, then
$I_q$ monotonously decreases when $q$ increases. 

The R\'enyi entropies are an extension of the Shannon information $H$.
In fact, $H$ is obtained in the limit $q\rightarrow 1$:
\begin{equation}
H=I_1=\lim_{q\rightarrow 1} I_q=-\sum_{i=1}^{N}p_i\log{p_i},
\end{equation}
where the constant $K$ of Eq. (\ref{eq0}) is considered to be the unity.
The disequilibrium $D$ is also related with $I_2=-\log\left(\sum_{i=1}^N p_i^2\right)$. 
We have that 
\begin{equation}
D=\sum_{i=1}^N p_i^2-{1\over N}=e^{-I_2}-{1\over N},
\end{equation}
then the LMC complexity is
\begin{equation}
C=H\cdot D=I_1\cdot\left(e^{-I_2}-{1\over N}\right).
\end{equation}
The behavior of $C$ in the neighborhood of $H_{max}$ takes the form 
\begin{equation}
C\sim {1\over N}(\log^2N-I_1I_2),
\end{equation}
The obvious generalization of the R\'enyi entropies for a normalized continuous
distribution $p(x)$ is
\begin{equation}
I_q={1\over 1-q}\log\int [p(x)]^q dx.
\end{equation}
Hence, 
\begin{eqnarray}
H & = & I_1=-\int p(x)\log p(x) dx, \\
D & = & e^{-I_2}=\int [p(x)]^2 dx.
\end{eqnarray}
The dependence of $\hat{C}=e^H\cdot D$ 
with $I_1$ and $I_2$ yields
\begin{equation}
\log\hat{C}= (I_1-I_2).
\end{equation}
This indicates that a family of different indicators could derive from
the differences established among R\'enyi entropies 
with different $q$-indices \cite{lopezruiz05}.
Let us remark at this point the coincidence of the indicator $\log\hat{C}$
with the quantity $S_{str}$ introduced by Varga and Pipek as a meaningful
parameter to characterize the shape of a distribution. 
They apply this formalism to the Husimi representation, i.e., 
to the projection of wave functions onto the coherent state basis \cite{varga}.
A further generalization of the LMC complexity measure as function of the R\'enyi 
entropies has been introduced in Ref. \cite{lopez09}.

The invariance of $\hat{C}$ under rescaling transformations implies
that this magnitude is conserved in many different processes.
For instance, the initial Gaussian-like distribution will continue to be 
Gaussian in a classical diffusion process. Then $\hat{C}$ is constant 
in time: ${d\hat{C}\over dt} = 0$, and we have:
\begin{equation}
{dI_1\over dt} = {dI_2\over dt}.
\end{equation} 
The equal losing rate of $I_1$ and $I_2$, i.e., the synchronization
of both quantities, is the cost to be paid in order
to maintain the shape of the distribution associated to the system and, hence, 
all its statistical properties will remain unchanged during its time evolution.

\section{Some Applications}

If by complexity it is to be understood that property present in all systems 
attached under the epigraph of `complex systems', 
this property should be reasonably quantified 
by the measures proposed in the different branches of knowledge.
In our case, the main advantage of LMC complexity is its generality and the fact that 
it is operationally simple and do not
require a big amount of calculations \cite{perakh04}. This advantage has been
worked out in different examples, such as the study of the time evolution 
of $C$ for a simplified model of an isolated gas, the ``tetrahedral gas" \cite{calbet01}
or also in the case of a more realistic gas of particles \cite{calbet07,calbet09},
the slight modification of $C$ as an effective method by which the
complexity in hydrological systems can be identified \cite{guozhang98}, 
the attempt of generalize $C$ in a family of simple complexity 
measures \cite{shiner99,martin03,lamberti04}, some statistical features
of the behavior of $C$ for DNA sequences \cite{zuguo00} or earthquake magnitude
time series \cite{lovallo05},
some wavelet-based informational tools used to analyze the 
brain electrical activity in epilectic episodes in the plane of 
coordinates $(H,C)$ \cite{rosso03}, a method to discern complexity
in two-dimensional patterns \cite{sanchez05-1}
or some calculations done on quantum 
systems \cite{panos05,sanudo08,montgomery08,kowalski09,lopez2009+}.
As an example, we show in the next subsections some straightforward 
calculation of the LMC complexity \cite{lopez01}.

\subsection{Canonical ensemble}

Each physical situation is closely related to a specific
distribution of microscopic states. Thus, an isolated
system presents equipartition, by hypothesis: the microstates
compatible with a macroscopic situation are equiprobable \cite{huang}.
The system is said to be in equilibrium.
For a system surrounded by a heat reservoir the probability of
the microstates associated to the thermal equilibrium follow the
Boltzmann distribution.
Let us try to analyze the behavior of $C$ in an ideal gas
in thermal equilibrium. In this case the probability $p_i$
of each accesible state is given by the Boltzmann distribution:
\begin{eqnarray}
p_i & = & \frac{e^{-\beta E_i}}{Q_N}, \\
Q_N & = & \int e^{-\beta E(p,q)} \frac{d^{3N}p d^{3N}q}{N! h^{3N}} =
e^{-\beta A(V,T)},
\end{eqnarray}
where $Q_N$ is the partition function of the canonical ensemble,
$\beta=1/\kappa T$ with $\kappa$ the Boltzmann constant and $T$ the
temperature, $V$ the volume, N the number of particles,
$E(p,q)$ the hamiltonian of the system, $h$ is the Planck constant
and $A(V,T)$ the Helmholtz potential.

Calculation of $H$ and $D$ gives us:
\begin{eqnarray}
H(V,T) & = & ( 1+T\frac{\partial}{\partial T})
\left( \kappa\log Q_N \right)  =   S(V,T), \\
D(V,T) & = & e^{2\beta \; \left[ A(V,T) - A(V,T/2) \right]}.
\end{eqnarray}
Note that Shannon information $H$ coincides with the thermodynamic
entropy $S$ when $K$ is identified with $\kappa$. If a system
verifies the relation $U=C_v T$ ($U$ the internal energy, $C_v$ the
specific heat) the complexity takes the form:
\begin{equation}
C(V,T) \sim cte(V) \cdot S(V,T) e^{-S(V,T)/\kappa}
\end{equation}
that matches the intuitive function proposed in Figure \ref{figIntro0}.

\subsection{Gaussian and exponential distributions}

{\bf Gaussian distribution}: Suppose a continuum of states represented
by the $x$ variable whose probability density $p(x)$ is given by the
normal distribution of variance $\sigma$:
\begin{equation}
p(x)=\frac{1}{\sigma \sqrt{2\pi}}\exp\left(-\frac{x^2}{2\sigma^2}\right).
\end{equation}
After calculating $H$ and $D$, the expression for $C$ is the following: 
\begin{equation}
C_g  =  H \cdot D \;\; = \;\; \frac {K}{2\sigma\sqrt{\pi}}\left(
\frac {1}{2}+\log(\sigma \sqrt{2\pi})\right).
\label{eq-gss}
\end{equation}

If we impose the additional condition $H\geq 0$, then
$\sigma\geq\sigma_{min}=(2\pi e)^{-1/2}$. The highest complexity is reached
for a determined width: $\bar{\sigma}=\sqrt{(e/2\pi)}$.

{\bf Exponencial distribution}: Consider an exponencial
distribution of variance $\gamma$:
\begin{equation}
p(x)=\left\{\begin{array}{ll}
\frac{1}{\gamma}e^{-x/\gamma} & x>0, \\
0                             & x<0.
\end{array} \right.
\end{equation}

The same calculation gives us:
\begin{equation}
C_e  =  \frac{K}{2\gamma}(1+\log\gamma),
\end{equation}
with the condition $H\geq 0$ imposing $\gamma\geq\gamma_{min}=e^{-1}$. 
The highest complexity corresponds in this case to $\bar{\gamma}=1$.

Remark that for the same width than a Gaussian distribution
($\sigma=\gamma$), the exponential
distribution presents a higher complexity ($C_e/C_g \sim 1.4$).

\subsection{Complexity in a two-level laser model}

One step further, combining the results obtained in the former sections,
is now done. We calculate LMC complexity for an unrealistic and
simplified model of laser \cite{svelto}.

Let us suppose a laser of two levels of energy: $E_1=0$ and $E_2=\epsilon$,
with $N_1$ atoms in the first level and $N_2$ atoms in the second level,
and the condition $N_1+N_2=N$ (the total number of atoms).
Our aim is to sketch the statistics of this model and
to introduce the results of photon counting \cite{arecchi}
that produces an asymmetric behavior of $C$ as function of
the population inversion $\eta =N_2/N$. In the range
$\eta\in (0,1/2)$ spontaneous and stimulated emission can take place,
but only in the range $\eta\in (1/2,1)$ the condition to have
lasing action is reached, because the population must be, at least,
inverted, $\eta>1/2$.

The entropy $S$ of this system vanishes when $N_1$ or $N_2$ is zero.
Moreover, $S$ must be homegenous of first order in the extensive
variable $N$ \cite{callen}.
For the sake of simplicity we approach $S$ by the first term in the
Taylor expansion:
\begin{equation}
S\sim \kappa \frac{N_1N_2}{N}= \kappa N\eta (1-\eta).
\end{equation}
The internal energy is $U=N_2\epsilon =\epsilon N\eta$ and the statistical
temperature is:
\begin{equation}
T=\left(\frac{\partial S}{\partial U}\right)_N^{-1}=
\frac{\epsilon}{\kappa}\frac{1}{(1-2\eta)}.
\end{equation}
Note that for $\eta>1/2$ the temperature is negative as corresponds
to the stimulated emission regime dominating the actual laser action.

We are now interested in introducing qualitatively
the results of laser photon counting in the calculation of
LMC complexity. It was reported in \cite{arecchi} that the
photo-electron distribution of laser field appears to be poissonian.
In the continuous limit the Poisson distribution is
approached by the normal distribution \cite{distri}.
The width $\sigma$ of this energy
distribution in the canonical ensemble is proportional to the statistical
temperature of the system. Thus, for a switched on laser in the regime
$\eta\in [1/2,1]$, the width of the gaussian energy distribution
can be fitted by choosing $\sigma\sim -T \sim 1/(2\eta-1)$
(recall that $T<0$ in this case).
The range of variation of $\sigma$ is
$[\sigma_{\infty},\sigma_{min}]=[\infty, (2\pi e)^{-1/2}]$.
Then we obtain:
\begin{equation}
\sigma \sim\frac{(2\pi e)^{-1/2}}{2\eta-1}.
\label{eqsigma}
\end{equation}
By replacing this expression in Eq. (\ref{eq-gss}),
and rescaling by a factor proportional
to entropy, $S\sim\kappa N$, (in order to give to it the correct
order of magnitude), LMC complexity for a population inversion
in the range $\eta\in [1/2,1]$ is reobtained:
\begin{equation}
C_{laser}\simeq \kappa N \cdot (1-2\eta)\log(2\eta-1).
\label{eqlaser}
\end{equation}
We can consider at this level of discussion
$C_{laser}=0$ for $\eta<1/2$.
Regarding the behavior of this function, it is worth noticing the value
$\eta_2 \simeq 0.68$ where the laser presents the highest complexity.
By following theses ideas, if the width, $\sigma$, of the
experimental photo-electron distribution of laser field
is measured, the population inversion parameter, $\eta$,
would be given by Eq. (\ref{eqsigma}). In a next step, 
the LMC complexity of the laser system would be obtained by Eq. (\ref{eqlaser}).

It is necessary to remark that a model helps us to approach 
the reality and provides invaluable
guidance in the goal of a finer understanding of a physical
phenomenon. From this point of view the present calculation evidently
only tries to enlighten the problem of calculating the LMC complexity
of a physical system via an unrealistic but simplified model.

\section{Conclusions}

A definition of complexity (LMC complexity) based on a probabilistic description
of physical systems has been explained. This definition contains basically an
interplay between the {\it information} contained in the system and the
{\it distance to equipartition} of the probability distribution representing
the system. Besides giving the main features of an intuitive notion of
complexity, we show that it allows to successfully discern situations
considered as complex in systems of a very general interest.
Also, its relationship with the Shannon information and the generalized R\'enyi entropies 
has been shown to be explicit. Moreover it has been possible to establish the
decrease of this magnitude when a general system evolves from a near-equilibrium 
situation to the equipartition. 

From a practical point of view, we are convinced that this statistical complexity measure 
provides a useful way of thinking \cite{hawking00} and it can help in the future to gain 
more insight on the physical grounds of models with potential biological interest.

%%%%%%%%%%%%%%%%%%%%%%%% referenc.tex %%%%%%%%%%%%%%%%%%%%%%%%%%%%%%
% sample references
% %
% Use this file as a template for your own input.
%
%%%%%%%%%%%%%%%%%%%%%%%% Springer-Verlag %%%%%%%%%%%%%%%%%%%%%%%%%%
%
% BibTeX users please use
% \bibliographystyle{}
% \bibliography{}

\end{document}